\documentclass[letter]{jpsj3}
\usepackage{txfonts}
\usepackage{color}
\usepackage{bm}
\usepackage{url}

\newcommand{\newblock{ }}

\title{Phase Diagrams of Three-Dimensional Anderson and Quantum Percolation Models
using Deep Three-Dimensional Convolutional Neural Network}

\author{Tomohiro Mano$^1$ and Tomi Ohtsuki$^1 $\thanks{ohtsuki@sophia.ac.jp}}
\inst{
$^1$Physics Division, Sophia University, Chiyoda-ku, Tokyo 102-8554, Japan} %

\abst{
The three-dimensional Anderson model is a well-studied model of disordered 
electron systems that shows the delocalization--localization transition.
As in our previous papers on two- and three-dimensional (2D, 3D) quantum phase transitions
[J. Phys. Soc. Jpn. {\bf 85}, 123706 (2016), {\bf 86}, 044708 (2017)],
we used an image recognition algorithm based on a multilayered convolutional neural network. However, in contrast to previous papers in which 2D image recognition was used,
we applied 3D image recognition to analyze entire 3D wave functions.
We show that a full phase diagram of the disorder-energy plane is obtained once the 3D convolutional neural network has been trained at the band center.
We further demonstrate that the full phase diagram for 3D quantum bond and site percolations
can be drawn by training the 3D Anderson model at the band center.
}

\begin{document}
\maketitle


{\it Introduction.}--
Applying machine learning methods to solve problems
in condensed matter physics has proven to be successful.
Ising and spin ice models\cite{Carrasquilla17,Tanaka17},
low dimensional topological systems,\cite{Zhang16,Zhang17}
strongly correlated systems,\cite{Carleo16,Broecker16,Chng16,Li16,Nieuwenburg17,Huang17,Schindler17,Saito17,Saito17b,Fujita17}
as well as random two- and three-dimensional (2D, 3D) topological and non-topological systems,\cite{Tomoki16,Tomoki17,Yoshioka17}
have been studied using machine learning.

In previous papers,\cite{Tomoki16,Tomoki17} we presented studies of 2D and 3D random electron systems
via a multilayer convolutional neural network (CNN) approach called deep learning, and the features of quantum phase transitions
such as delocalization--localization transitions (the Anderson metal--insulator 
transition) and topological--nontopological insulator transitions were
shown to be captured.
We diagonalized the Hamiltonian, obtained the eigenfunctions $\Psi_\nu(\bm{r})$, and trained the neural network by
feeding the electron density $|\Psi_\nu(\bm{r})|^2$ for a specific quantum phase.
We used 2D image recognition, and for 3D systems, integration over one direction was
performed to reduce the 3D electron density to 2D.
The advantage of reducing the electron density from 3D to 2D is that we can then use standard 2D image recognition.
The drawback is that we lose the information of electron density distribution along one direction.

In this Letter, we apply 3D image recognition via deep 3D CNN to
analyze a random 3D electron system.
The Anderson model for delocalization--localization transition\cite{Anderson58}
and the 3D quantum bond and site percolation models were studied.
We show that the full phase diagram of the disorder-energy ($W$--$E$) plane is available once the 3D CNN has been trained at the band center, i.e., along $E=0$ (see arrows in Fig.~\ref{fig:AT}(b)).
We further demonstrate that the two-dimensional phase diagrams
(see Fig. ~\ref{fig:QPPhaseDiagram}) for the 3D quantum percolation model,
both bond and site types,
can be drawn by training the 3D Anderson model at the band center, without learning the features of quantum percolations.

{\it Models and method.}--
We consider the following 3D Anderson Hamiltonian,\cite{Anderson58}
\begin{equation}
\label{eq:3dandersonHamiltonian}
H=\sum_{\bm{x}}
v_{\bm{x}} c_{\bm{x}}^\dagger c_{\bm{x}}+
\sum_{\langle \bm{x},\bm{x}'\rangle} c_{\bm{x}}^\dagger c_{{\bm{x}}'}\,,
\end{equation}
where $c_{\bm{x}}^\dagger$ ($c_{\bm{x}}$) denotes the creation (annihilation)
operator of an electron at a site $\bm{x}=(x,y,z)$ that forms a simple cubic lattice,
and $v_{\bm{x}}$ denotes the random potential at $\bm{x}$.
$\langle \cdots \rangle$ indicates the nearest-neighbor hopping.
A box distribution is assumed with
each $v_{\bm{x}}$ uniformly and independently distributed at the interval
$[-W/2,W/2]$.
At energy $E=0$, i.e., at the center of the band, the wave functions are delocalized
when $W<W_c$ and the system is a diffusive metal.  For $W>W_c$, the wave functions
are exponentially localized and the system is an Anderson insulator.
Here, the critical disorder $W_c$ at $E=0$ is estimated to be $16.54\pm 0.01$\cite{Slevin14} by the finite size
scaling analysis of the Lyapunov exponent calculated from the transfer matrix
method.\cite{MacKinnon81,MacKinnon83,Pichard81,Kramer93}

Next, we consider the 3D quantum bond and site percolation models described by the following Hamiltonian,\cite{Avishai92,Berkovits96,Kaneko99,Ujfalusi14}
where electrons are allowed to move on connected sites of a classical percolation,\cite{StaufferBook,
Chakrabarti09,Saberi15}
\begin{equation}
\label{eq:percolationHamiltonian}
H=\sum_{\bm{x}}
v_{\bm{x}} c_{\bm{x}}^\dagger c_{\bm{x}} +
\sum_{\langle \bm{x},\bm{x}'\rangle} 
t_{\bm{x}\bm{x}'} c_{\bm{x}}^\dagger c_{{\bm{x}}'}\,.
\end{equation}
The hopping parameter $t_{\bm{x}\bm{x}'}$ is defined as
\begin{equation}
\label{eq:transfer}
t_{\bm{x}\bm{x}'}=\left\{
\begin{array}{ll}
   1   &  \left({\rm for\;connected\;bond}\right) \,\\
   0   &  \left({\rm for\;disconnected\;bond}\right)\,,
\end{array}
\right.
\end{equation}
where the energy unit is the transfer energy between connected sites.
In the case of bond percolation, bonds are randomly connected with a probability $p_{\rm{B}}$.
In the case of site percolation, each site is filled with a probability $p_{\rm{S}}$, and bonds are
connected only if the sites on both sides of the bond are filled. 
For each realization of bond and site percolations, we identify the maximally connected cluster
with a depth-first search, and analyze only the states in this cluster.

According to the studies on quantum percolation,\cite{Kirkpatrick72,Chayes86a,Schubert09,Ujfalusi14}
many strongly localized states, so-called molecular states,
appear at energies, e.g., $E=0, \pm 1, \pm \sqrt{2}$, when we set  $v_{\bm{x}}=0$.
These states are peculiar to the quantum percolation model, and are degenerate, resulting in strong peaks in the density of states.
Due to this degeneracy, any linear combination is possible, which results in difficulty in judging the delocalized/localized phases.
We therefore assume a very weak site random potential of the order of $10^{-3}$, namely, $v_{\bm{x}}\in [-10^{-3}/2, 10^{-3}/2]$, and thereby
lift the degeneracy.

Random numbers were generated by the Mersenne Twister algorithm,\cite{Matsumoto98} MT2203.
We imposed the periodic boundary condition, and the maximum modulus of the eigenfunction was
shifted to the center of the system to improve the accuracy of the 3D image recognition.
For all calculations,
we considered a $40\times 40\times 40$ lattice, and diagonalized the Hamiltonian.
For training data and the data for Fig.~\ref{fig:AT}(a),
we used the sparse matrix diagonalization Intel MKL/FEAST,\cite{Polizzi09} since we restricted the energy range to around $E=0$.
For drawing the phase diagrams, we used the standard Linear Algebra package LAPACK
\cite{lapack}
for diagonalization, and obtained all the eigenvalues and eigenvectors.

\begin{figure}[htb]
  \begin{center}
\includegraphics[angle=0,width=0.28\textwidth]{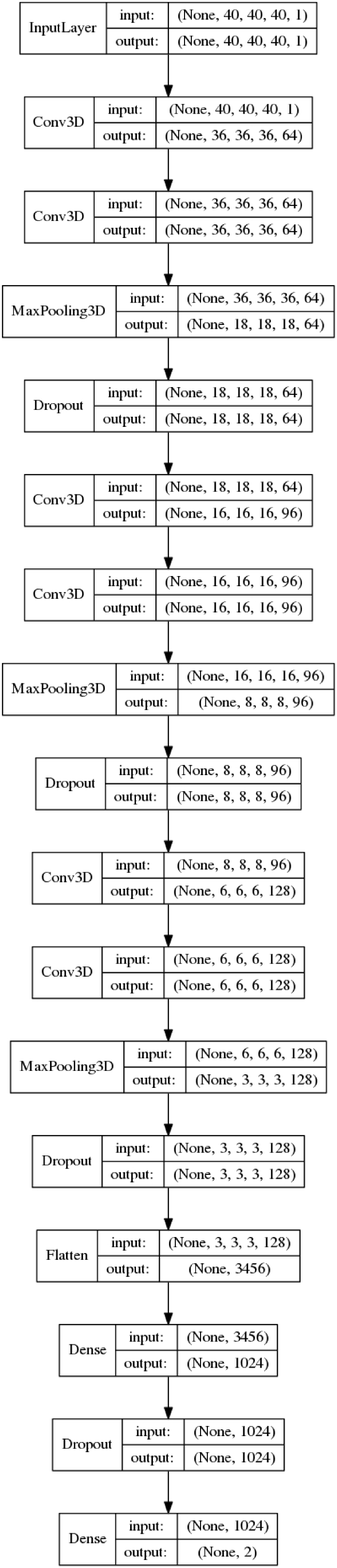}
 \caption{Schematic of the deep 3D CNN structure, where the neural network parameters are indicated.
 6 convolutions have been performed, with the 1st and 2nd convolution $5\times 5\times 5$ with stride 1, and
 the other convolutions, $3\times 3\times 3$ with stride 1.
 For the 2nd, 4th, and 6th convolutions, zero padding is applied so that the input and the output image sizes are the same.
 The activation function ReLU was performed after each convolution layer as well as the densely-connected 
 layer, except for the last layer where softmax was used.}
\label{fig:CNNSchematic}
\end{center}
\end{figure}

\begin{figure}[htb]
  \begin{center}
     \begin{tabular}{cc}    
 
 \begin{minipage}{0.45\hsize}
  \begin{center}
   \includegraphics[angle=0,width=\textwidth]{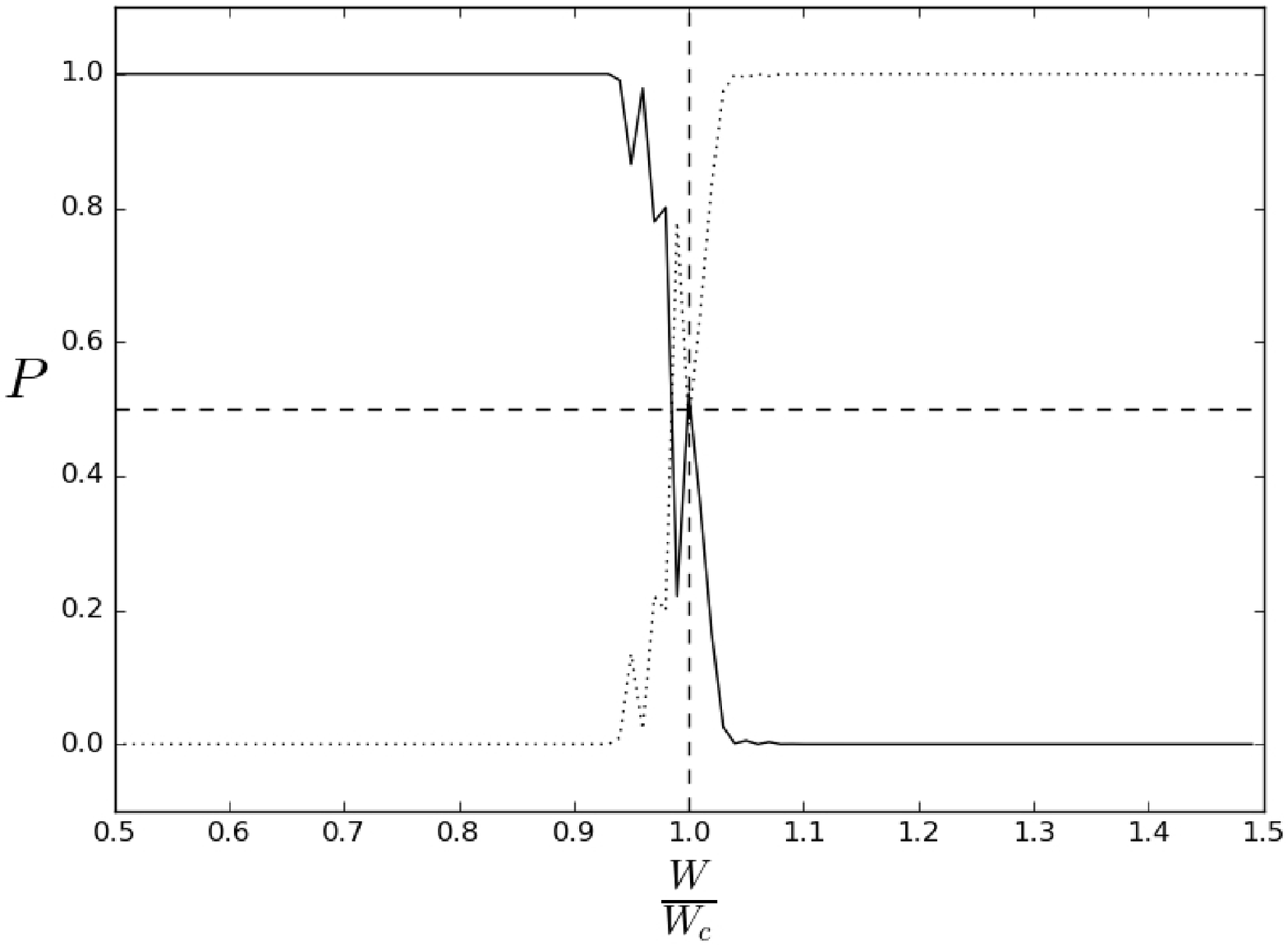}
      \hspace{1.6cm} (a)
     \end{center}
 \end{minipage}
 \begin{minipage}{0.55\hsize}
  \begin{center}
  \includegraphics[angle=0,width=\textwidth]{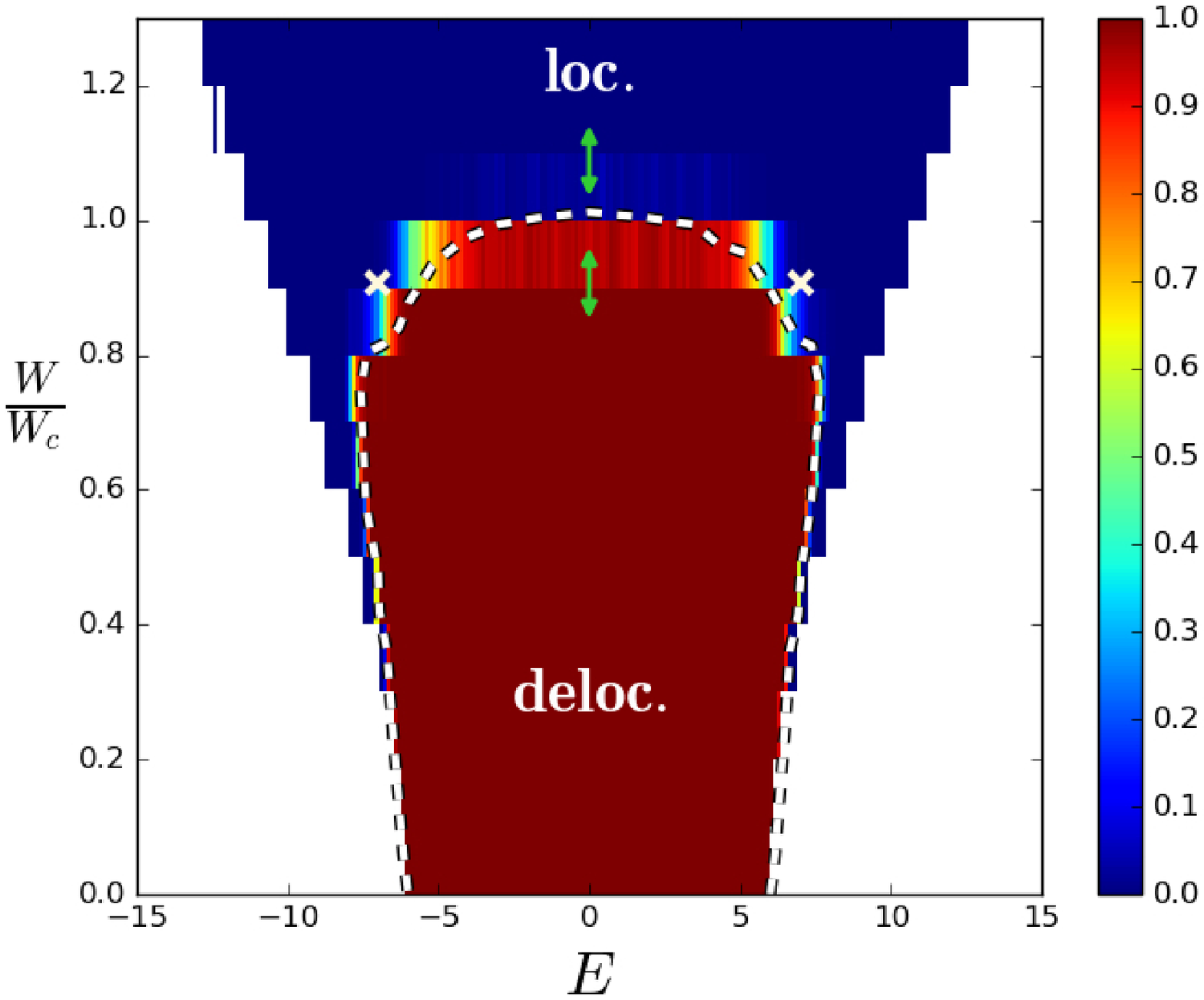}
     \hspace{1.6cm} (b)
     \end{center}
 \end{minipage}
  \end{tabular}

 \caption{(Color online) The probability that the states are delocalized/localized for the Anderson model.
 $W_c=16.54$ is the critical disorder for $E=0$.  The training regions, which are one-dimensional line segments,
 are indicated by the arrows.
 (a) $P_\mathrm{deloc}$ (solid line) and $P_\mathrm{loc}$ (dotted line) when the energy $E$ is close to the band center.
 50\% probability as well as $W=W_c$ is indicated as dashed lines as a guide to the eye. (b) Color map of the intensity
 $P_\mathrm{deloc}$ for all the energy spectrum.
In both figures, an average over 5 samples was performed.
Dashed line is from ref.~\cite{Schubert05}, while the crosses ($\times$) are from ref.~\cite{Queiroz01}.}
\label{fig:AT}
\end{center}
\end{figure}

\begin{figure}[htb]
  \begin{center}
 
 \begin{minipage}{0.45\hsize}
  \begin{center}
   \includegraphics[angle=0,width=\textwidth]{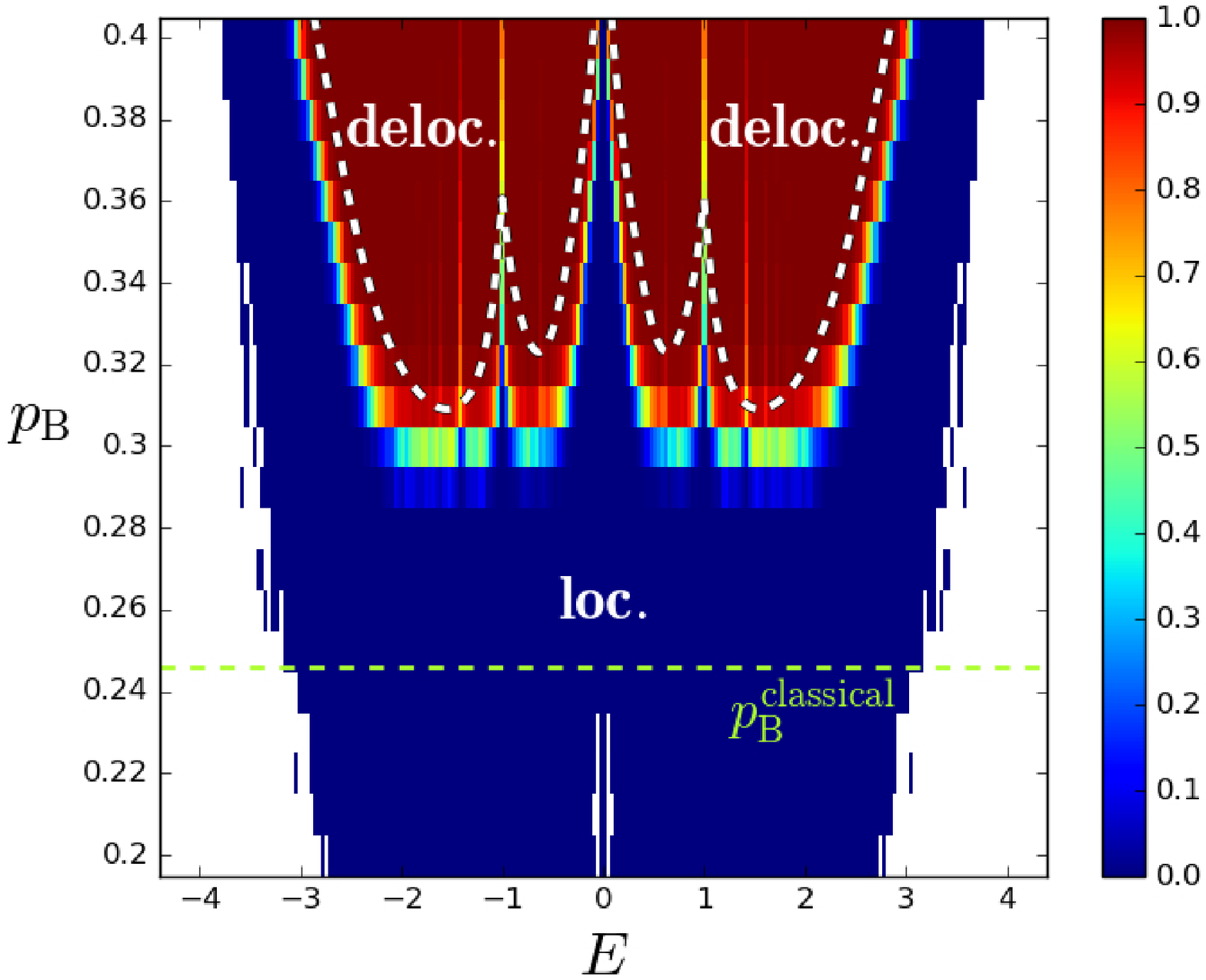}
      \hspace{1.6cm} (a)
     \end{center}
 \end{minipage}
 \begin{minipage}{0.45\hsize}
  \begin{center}
  \includegraphics[angle=0,width=\textwidth]{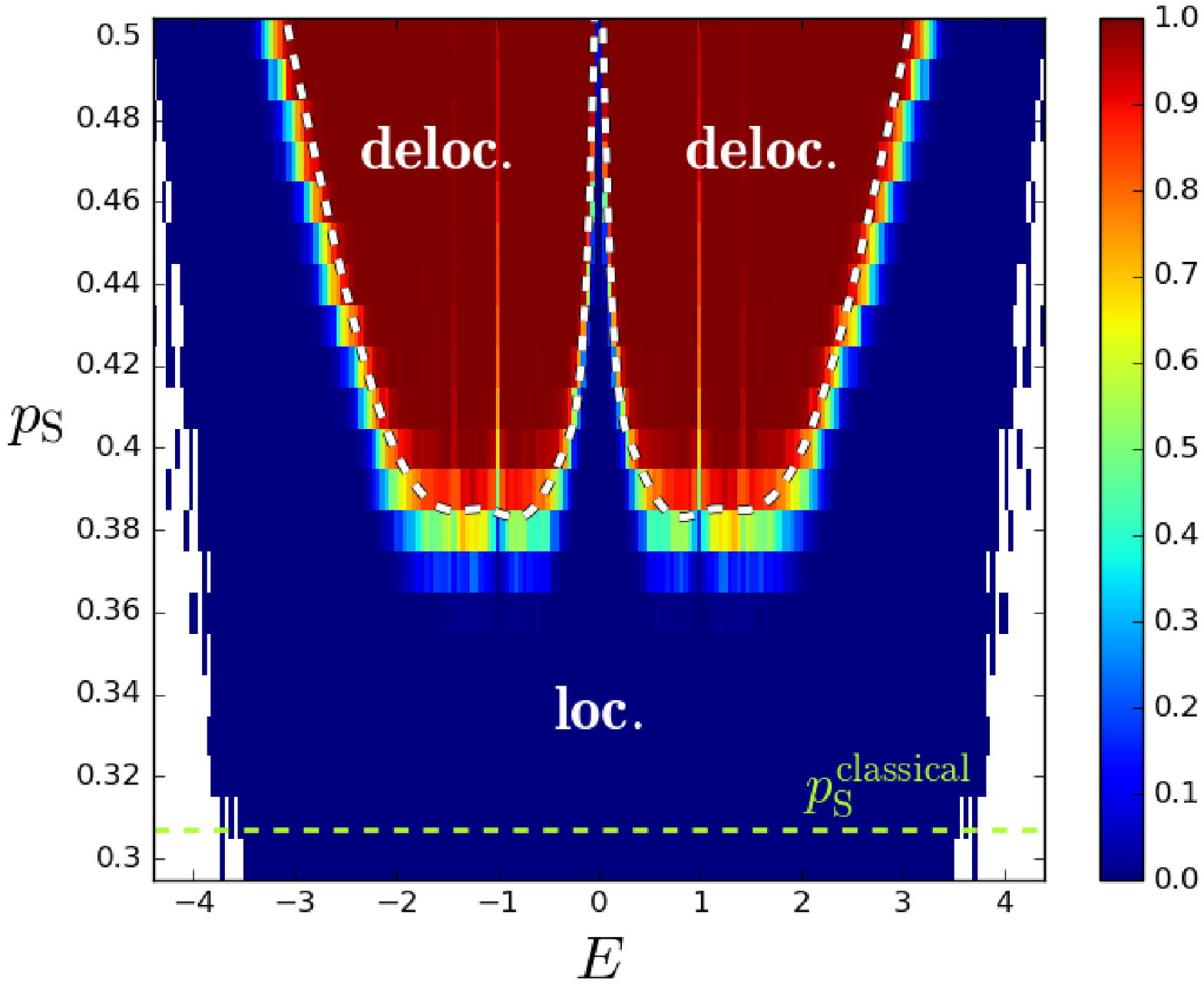}
      \hspace{1.6cm} (b)
     \end{center}
 \end{minipage}

 \caption{(Color online) Phase diagram of the (a) bond and (b) site quantum percolations.
The intensity $P_\mathrm{deloc}$ is shown.
Classical percolation thresholds for the bond percolation, $p_\mathrm{B}^\mathrm{classical}$, and 
for the site percolation, $p_\mathrm{S}^\mathrm{classical}$, are indicated by green dashed lines.
In both figures, the average over 5 samples has been performed.
The white dashed line in (a) is a spline interpolation from the data in ref.~\cite{Soukoulis92},
while that in (b) is the spline interpolation from the data in ref.~\cite{Ujfalusi14}.}
\label{fig:QPPhaseDiagram}
\end{center}
\end{figure}

To calculate the probabilities of delocalized and localized phases, $P_\mathrm{deloc}$ and $P_\mathrm{loc}(=1- P_\mathrm{deloc})$ for a given eigenfunction,
we adopted a 3D CNN using Keras\cite{chollet2015keras} as frontend and TensorFlow as backend. NVIDIA GTX GEFORCE 1080 and 1080 Ti graphics cards were used for GPU calculation.
The configuration of our CNN and detailed parameters are shown schematically
 in Fig.~\ref{fig:CNNSchematic}. The input shape was $40\times 40\times 40$, corresponding to the system size of our numerical simulation.
No bias parameters are included in the weight parameters.
Through the hidden layers that consist of a convolution layer, max pooling layer, and densely-connected layer, our neural network outputs two real numbers, i.e., the probability of a delocalized (localized) phase
$P_\mathrm{deloc}\, (P_\mathrm{loc})$ in the case of the Anderson and quantum percolation models. 
The network utilizes the AdaDelta solver\cite{Zeiler12} as the stochastic gradient decent solver and rectified linear unit (ReLU) as its activation function except for the last layer, which is activated by the softmax function.\cite{Goodfellow16}
The loss function is defined by the cross entropy (which is {\it{categorical\_crossentropy}} in Keras),
which was minimized during the training.
To avoid overfitting, dropout processes, which randomly drop half of the inputs, were implemented in some layers (see Fig.~\ref{fig:CNNSchematic}). 
Before training, we randomly chose 10\% of the training data as a validation data set.

We then trained our neural network based on data from the 3D Anderson model, and determined
 the network weight parameters, which captured the features of the delocalized/localized phase. 
 For training data, we restricted the energy region to the vicinity of the band center, $E=0$, in the Anderson model.
The main purpose of this work was to obtain global delocalization--localization phase diagrams of 3D random electron systems, and
we assumed that the critical disorder at $E=0$, $W_c=16.54$, was known in advance.
As the training data of the delocalized phase, we prepared 4000 eigenstates in the range, $W\in [14.0,16.0]$. As the training data for the localized phase, we prepared 4000 eigenstates in the range, $W\in [17.0,19.0]$.
These training regions are indicated by arrows in Fig.~\ref{fig:AT}(b).

To confirm that the network sufficiently captures the features of the eigenstates,
we prepared test data from the Anderson model for various $W$ and $E$, and determined the phase diagram
in $W$--$E$ parameter space and then compared it with those obtained by other researchers.\cite{Bulka87,Queiroz01,Schubert05}
Once the results were confirmed, we applied this trained CNN to bond and site quantum percolation models,
the phase diagrams of which are less well-known.
5 independent samples were prepared for the test data from the Anderson and quantum percolation models,
and the output probabilities are averaged over the 5 samples.

{\it Result for Anderson model.--}
In Fig.~\ref{fig:AT}(a), we show the probabilities $P_\mathrm{deloc}$ and $P_\mathrm{loc}$ at $E=0$ as functions of $W$. The test data 
consisted of $100\times 5$ eigenfunctions closest to $E=0$:
100 values of $W\in[\frac{1}{2}W_c\,,\,\frac{3}{2}W_c]$, and 5 independent samples.
The average over 5 samples was taken as in Fig.~1 of reference.\cite{Tomoki17}
From this figure, the Anderson metal--insulator transition from delocalized (metal) phase
to localized (insulator) phase has been observed around $W_c$,
from which we can confirm that the CNN correctly detected the Anderson transition. 

We then prepared eigenfunctions all over the energy spectra with varying $W$,
and let the CNN determine the phase. The result is shown in Fig.~\ref{fig:AT}(b)
where the intensity $P_\mathrm{deloc}$
is plotted as a color map,
which is in good agreement with the previous results from the transfer matrix.\cite{Bulka87,Queiroz01} 
The flat phase boundary around $W_c, |E|\lesssim 6$, together with the rapidly changing critical
disorder outside the band edge $|E|=6$ in the absence of randomness, are reproduced.

For relatively small disorder, the CNN judges that eigenstates near the band edge are localized.
With increasing $W$ but fixed $E$ around $E=7$, the localized
states near the band edge are delocalized and then localized again,
a re-entrant phenomenon known for the Anderson model with the box distributed site energies.\cite{Bulka87,Queiroz01,Schubert05} 

Note that the states near the band edge with small disorders look rather like bound states
trapped by the potential fluctuation, and smoothly change in space.
It is interesting that the CNN trained only around $E=0$, where the random quantum interference was strong and the wave function amplitudes were rapidly fluctuating, 
correctly judged the smooth bound states as localized.


{\it Application to quantum percolation.--}
Figs.~\ref{fig:QPPhaseDiagram}(a) and (b) show the phase diagrams of the bond and site quantum percolation models, respectively.
Eigen energies of the maximal clusters are plotted on the horizontal axis, while the probability $p_\mathrm{B}$ or $p_\mathrm{S}$
is plotted on the vertical axis.  All over the energy spectra,
the quantum percolation transition occurs well above the classical percolation
threshold, $p_\mathrm{B}^\mathrm{classical}\approx 0.2488$ and $p_\mathrm{S}^\mathrm{classical}\approx 0.3116$.\cite{Sur76,Wang13}.
The energy dependence of the quantum site percolation threshold is consistent with the finite
size scaling analysis of multifractality\cite{alberto10,alberto11,Ujfalusi15} by Ujfalusi and Varga,\cite{Ujfalusi14}
where a strong energy dependence is observed near $E=0$ and near the band edge.
\cite{Soukoulis92,Kusy97,Travenec08}

In the quantum percolation models, even on the maximal cluster, there appear
energetically degenerate molecular states,
the amplitudes of which are localized on finite sites, while being exactly zero on other sites.\cite{Kirkpatrick72,Chayes86a}
As stated while explaining the model, we lifted the degeneracy by introducing a very weak site randomness.
The molecular states can coexist with other states, but their numbers are much smaller than
other non-molecular states except at specific energies $E=0, \pm 1, \pm \sqrt{2}, \cdots$
where the shapes of molecular states are relatively simple.
Because we average the probability $P_\mathrm{deloc}$ over a finite energy region,
the phase is judged to be delocalized (metal) if the other non-molecular states are delocalized.

{\it Summary and concluding remarks.--}
In this paper, we applied 3D image recognition to draw the global phase diagrams of 3D random electron systems.
We have shown that once the 3D CNN is trained in a small region of the Anderson transition, the global phase
diagram can be obtained by using this training.  Furthermore, the CNN trained for the Anderson model at $E=0$
allows us to draw the phase diagrams of the quantum bond and site percolation models.
Thus, the present work has opened a way to use 3D deep CNN for condensed matter physics.
Application to strongly correlated 3D electron systems with disorder\cite{Shinaoka09,Shinaoka10,Harashima14} 
is one of the interesting problems that could benefit from this approach.
Another interesting application is to distinguish rare event states in Weyl/Dirac semimetals\cite{Nandkishore14,Pixley16}
 from other types of states,
which is especially important in discussing the Weyl/Dirac semimetal to metal transition.
\cite{Kobayashi14,Ominato14,Syzranov15,Liu16}

The present method can be extended easily to analyze other quantum percolation problems,
such as systems with magnetic fields or spin--orbit interactions,
 i.e., systems belonging to different symmetry classes.\cite{Dyson61,Dyson62,Altland97,Evers08}
In these systems, the quantum percolation threshold is expected to be lowered due to the
 breaking of the time reversal or spin--rotational symmetries.\cite{Kaneko99}
How the energy-dependent quantum percolation threshold is changed by the breaking of
the symmetries is an interesting problem that remains to be studied in the future.
 
 Let us conclude this Letter by comparing the present method with the transfer matrix
 method,\cite{MacKinnon81,MacKinnon83,Pichard81,Kramer93,Slevin14} and stating that
the present method and the transfer matrix method are complementary.
Due to the zero connections between sites that appear randomly, the transfer matrix analysis is
not applicable to the quantum percolation problem.
This situation is
similar to topological insulators with a specific set of parameters.\cite{Tomoki17}
The method presented here is, in this sense, more general than the transfer matrix method.
In addition, we can draw the phase diagram in the $W$--$E$ plane easily using the present method,
while the transfer matrix method requires a one-dimensional parameter change,
hence drawing the global $W$--$E$ phase diagram is more numerically demanding.
On the other hand, the transfer matrix method gives a more precise estimate of the critical point than the present method.
The estimate of the critical exponent is also possible in the finite size scaling
analysis of the Lyapunov exponent obtained by the transfer matrix method,
which is so far impossible using the present method.
A possible way to estimate the critical exponent by the present method is to change the
system size $L$ and observe how the quantities, such as probabilities of delocalized/localized states and/or convergence time
vary with $L$.  The independent estimate of critical disorder $W_c$, with the aid of unsupervised training,
is also an important topic for the future.

\begin{acknowledgments}
The authors would like to thank Dr. Tomoki Ohtsuki, Dr. Koji Kobayashi, Prof. Imre Varga, and Prof. Keith Slevin for their useful comments.
This work was partly supported by JSPS KAKENHI Grant No. JP15H03700 and JP17K18763.
\end{acknowledgments}

\newpage

\bibliography{manoJPSJ17}

\end{document}